\documentclass[preprint,5p]{elsarticle}
\usepackage{graphicx}
\usepackage{amsmath,amsfonts}
\usepackage{dcolumn}
\usepackage{bm}
\usepackage{slashed}
\usepackage{nicefrac}
\usepackage{epstopdf}
\usepackage{color}
\def\bra{\langle}
\def\ket{\rangle}

\def\cL{\mathcal{L}}

\def\I{\mathbb{I}}

\def\Tr{\mathrm{Tr}}

\def\tr{\mathrm{Tr}}
\begin{document}
\begin{frontmatter}
\title{Quantum reservoir networks based on decoherence-free subspaces}
\author[mgu]{V.V. Akshay}
\address[mgu]{School of Nanoscience and Nanotechnology, Mahatma Gandhi University, Priyadarsini Hills, Kottayam, Kerala, 686560, India}
\ead{vvakshay285@gmail.com}
\author[iki]{M.V. Altaisky} 
\address[iki]{Space Research Institute RAS, Profsoyuznaya 84/32, Moscow, 117997, Russia}
\ead{altaisky@cosmos.ru}
\author[misis]{N.E. Kaputkina}
\address[misis]{National University of Science and Technology 'MISIS', Moscow, 119049, Russia}
\ead{kaputkina.ne@misis.ru}
\begin{abstract}
We present the numerical simulation of a six-qubit 
quantum reservoir network with an output implemented on a 5-dimensional decoherence-free subspace (DFS), which classifies entangled and product states of the input quantum system fed into the reservoir during a finite learning time. The DFS are dynamically decoupled from the environment, hence evolve unitarily and are therefore decoherence-dissipation free. In principle, this approach could complement or coexist with active quantum error correction. Further, the realisation of effective hardware and readout mechanisms could open possibilities for energy-efficient quantum information processing systems, potentially extending to future quantum machine learning architectures with relaxed cooling constraints.
\end{abstract}
\end{frontmatter}
\section{Introduction \label{intro:sec}}
Quantum computing (QC) and artificial intelligence (AI) are among the most active branches of modern computer science. Both have earned their credibility through data-processing parallelism, allowing them to outperform conventional algorithmic approaches. However, both share the bottleneck of huge energy and hardware overheads for different reasons. These challenges make the creation of commercially effective AI systems and practical quantum information processing systems extremely problematic \cite{Lake2017} and motivate approaches that reduce resource demands, making ideas such as passive noise protection using decoherence-free subspaces highly relevant, along with traditional error correction routes. 

At the beginning of the quantum computational era, when Deutsch 
proposed a circuit-based quantum computer \cite{Deutch1985}, 
the main attention was paid to the computational efficiency 
of quantum algorithms. In this way, the Shor algorithm, being based on the quantum fast Fourier transform, 
can solve classically intractable factorization problems in polynomial time due to the inherent quantum parallelism of information processing  \cite{Shor1994}. For the same reason, the Grover database search algorithm finds an unknown record in a database 
in $\sqrt{n}$ steps only, instead of $n/2$ steps of a classical 
search \cite{Grover1997}.

The main practical problem for algorithmic circuit-based quantum 
computers remains to keep many qubits of a quantum data processing unit in a state of quantum superposition, which provides the quantum parallelism of data processing. The quantum 
superposition states are rather fragile and can be destroyed by 
merely small fluctuations of the environment. This is the {\em decoherence problem} \cite{Zeh1970,Guilini1996}. Because of decoherence,
most of the present quantum data processing systems operate at 
cryogenic temperatures. Examples include superconducting qubits and quantum dots, although optical and some solid-state systems can potentially operate at room temperature. 

Low operating temperature implies an expensive cooling system 
and limits scalability and portability of the technology, which are essential for compact
devices. The standard remedy for the computational errors caused by decoherence is quantum error-correcting codes (QECC),  implemented by a system of quantum gates, which can correct the erroneous qubits based on the redundant encoding processed by a given algorithm \cite{Shor1995ECC, Stean1996ECC}.

Surprisingly, the initial idea of quantum computers, as first proposed by R. Feynman \cite{Feynman1982}, was not algorithmic at all—it was a simulation of the physical system of interest by a smaller quantum system, i.e., a direct simulation. Modern AI systems, which consume enormous amounts of energy on classical hardware and await a future energy-efficient implementation, are not algorithmically prescribed in the same sense; they are based on neural networks. The idea of a neural network (NN) is a mathematical endeavor to describe the work of the brain, first formulated by Hebb \cite{Hebb1949}. In a general sense, a neural network is a massively parallel distributed processor made up of simple processing units (neurons) that has a natural propensity for storing experiential knowledge (in inter-neuron connections) and making it available for use \cite{Haykin2009}. In the case of a hardware implementation of an artificial neural network (ANN), the change of weights—connections between neurons—during the learning process is performed by learning algorithms implemented on classical computational hardware. In a physical or biological implementation, by contrast, the learning process is not governed by an externally imposed algorithm: the network evolves according to the laws of physics, and learning emerges from its natural dynamics.  

The reason for the tremendous effectiveness of the human brain in comparison to numerical computer algorithms may not be only due to the inherent 
parallelism of neural networks (which is also true for modern 
deep-learning systems \cite{CBH2015}), but also due to the quantum nature of data processing. Different authors suggest the idea that the human brain is a kind of quantum computer
\cite{Chav1970e, BE1992, HHT2002, AP2020, Kurian2025}. The main argument that precluded the scientific community from taking this idea seriously was Tegmark's estimation of 
decoherence time for biomolecular systems to be about 
$\tau_\mathrm{decoh} \propto 10^{-19}\div 10^{-13}s$, from ions to microtubules,
respectively \cite{Tegmark00}. However, these calculations assume thermal equilibrium and uncontrolled ion scattering in the system without explicitly accounting for sparse activation or the brain's structural complexity. Considerations such as these keep the nature of the biological brain an open question.

Higher coherence lifetimes in biological systems at elevated temperatures have been suggested previously \cite{Collini2010}. Whether analogous mechanisms operate in neural tissue is unclear. However, room-temperature quantum coherence is not to be ruled out a priori by thermodynamics alone. This motivates a narrower and more tractable idea of a controllable quantum system being engineered so that its output subspace is intrinsically insensitive to the environment.

The idea of quantum reservoir computing (QRC), which exploits the natural quantum dynamics of ensemble systems, was first proposed by Fujii and Nakajima in 2017 \cite{PhysRevApplied.8.024030}. Ghosh et al. \cite{Ghosh2019} later extended the idea to address quantum information processing. Recent QRC implementations include gate-based superconducting reservoirs, where non-unital amplitude damping supplies fading memory without active reset \cite{Monzani2024-gl}, Rydberg-atom and microwave-circuit reservoirs for signal processing and large-scale learning \cite{Dudas2023-nf}, and cavity-QED reservoirs based on Jaynes–Cummings and Tavis–Cummings atom–photon systems, where the shared cavity mode couples the atoms and continuous field measurement provides the readout \cite{Zhu2025-no} are some of them. These schemes exploit noise and dissipation as a computational resource, while a broad advantage over decoherence and dissipation still remains a challenge for these implementations. 

A mathematical model for avoiding dissipation has been 
proposed by several authors in the context of quantum computers 
\cite{ZR1997,LCW1998}.
It is based on symmetry considerations of the system-environment 
interaction and consists of encoding the information in 
decoherence-free subspaces (DFS)\cite{ZR1997}. DFS have been actively discussed in the context of circuit-based quantum computers but were first proposed for application in quantum 
neural networks in \cite{AK2025J}.

Here we consider the  model of a six-qubit 
reservoir-type quantum neural network, the output of which is 
formed by the projections of the six-qubit reservoir network state  
onto a 5-dimensional DFS. The dimension 5 turns out to be sufficient for fairly good classification of both two-qubit input states and squeezed input states in the Fock basis in two 
classes of entangled and product states, respectively.   

The remainder of this paper is organised as follows. In {\em Sec.}~\ref{dfs:sec} we recall the ideas 
of DFS, as was proposed in \cite{ZR1997}. In {\em Sec.}~\ref{qrc:sec} we outline the model of 
quantum reservoir network used for quantum data classification. {\em Sec.}~\ref{dc:sec} presents the 
dynamics of quantum reservoir network used for classification of input quantum states into 
entangled and product classes. The simulations are presented in Lindblad approximation for two 
different types of input states: (A) two-qubit teacher states; (B) two-mode squeezed teacher states.

\section{Decoherence-free subspaces \label{dfs:sec}}
The dynamics of any quantum information processing system (whatever it be a circuit-based computer, quantum neural network, or anything else) embedded into a noisy environment can be described by a Hamiltonian of the form 
\begin{equation}
	H = H_S + H_B + H_I,
\end{equation}
where $H_S$ is the Hamiltonian of the processing system itself, 
$H_B$ is the Hamiltonian of the environment, and $H_I$ is the Hamiltonian 
describing the interaction between the system and its environment.

The combined system (data processing unit and its environment) is 
a closed quantum system, and its quantum evolution should be unitary. However, we cannot be aware of all details of the environment 
dynamics and have to average over the states of environment 
using the theory of open quantum systems (OQS) \cite{BP2002}, 
i.e., we have to trace out the environment degrees of freedom 
to obtain the density matrix of the processing system itself.

In brief, the environment (B) works on the system (S) as a 
measuring device, which results in the decoherence -- decay of the off-diagonal terms of the system density matrix $\rho_S$ \cite{PSE1996}. The principle role in the decoherence effects 
(or their absence) is played by the system-environment interaction Hamiltonian $H_I$, or, more precisely, by its 
symmetries.

Suppose a quantum register is represented by a usual register 
of $N$ spin-bosons \cite{Leggett1987RMP} described by a Hamiltonian 
\begin{equation}
	H_S = \epsilon \sum_{i=1}^{N}\sigma_i^z ,
	\label{eq:HS}
\end{equation}
Here we assume identical qubits subject to a uniform field, so that a
single $\epsilon$ characterizes all qubits and $H_S=\epsilon S^z$ is
expressed through the collective spin operator.
Here $\sigma^z_i$ is the Pauli matrix acting on the $i$-th 
qubit, and the environment is represented by a heat bath of 
harmonic oscillators, described by the Hamiltonian 
\begin{equation}
	H_B = \sum_k \omega_k b^\dagger_k b_k,
\end{equation}
with the sum taken over all oscillator modes.

In the first-order approximation, the interaction Hamiltonian 
$H_I$ describes spin flips in the system register caused by the absorption/emission of oscillator modes from/to the heat bath:
\begin{equation}
	H_I = \sum_{i=1}^N \sum_k g_{ki}\sigma_i^+ b_k + f_{ki}\sigma_i^- b_k^\dagger + h_{ki} \sigma_i^z b_k + h.c.,
\end{equation}
where $\sigma^\alpha_i$ is the Pauli matrix acting on  $i$-th qubit.

If the coupling constants $g_{ki},f_{ki},h_{ki}$ {\em do not depend on the qubit number $i$}, one can sum up over all 
qubits and construct the full Hamiltonian, describing the interaction of the whole register with the heat 
bath:
\begin{equation}
	H_I = \sum_k g_k S^+ b_k + f_k S^- b_k^\dagger + h_k S^z b_k + h.c., \label{hi}
\end{equation}
$$\hbox{where\ }
S^\alpha = \sum_{i=1}^N \sigma^\alpha_i, \quad \alpha = \pm,z,
$$
are the total spin operators of the qubit register.

The independence of coupling constants $g_{ki},f_{ki},h_{ki}$
on the qubit number implies the wavelength of the environment oscillations is much bigger 
than the geometric size of the system $S$. This is known as the generalization of Dicke limit in quantum optics \cite{HL1973}.

Since each qubit in the register can be understood as a spin-one-half representation of $SU(2)$ group, i.e., is described 
by a two-component spinor $\begin{pmatrix}
	\uparrow \cr \downarrow
\end{pmatrix}$, for each {\em even} value of $N$, the state of the whole register, which is the product of individual states, 
contains singlets. As it is known from group representation theory \cite{Cornwell:book}, the decomposition 
of a direct product of representations into a sum of representations 
$$
D^{\otimes N}_\frac{1}{2} = \sum_{j\in J} n_j D_j,
$$
where $D_j$ is irreducible representation of $SU(2)$ group, corresponding to the 
total angular momentum value $j$ and having the dimension $(2j+1)$, with $n_j$ being the 
number of times representation $D_j$ appears in the sum.  
For the $SU(2)$ group this gives the Clebsch-Gordan series:
\begin{align}\nonumber 
	\nonumber D_{1/2}\otimes D_{1/2} &=& D_1 \oplus D_0, \\
	D_{1/2}^{\otimes 4}    &=& D_2 \oplus 3 D_1 \oplus 2 D_0, \label{cgs} \\
	\nonumber D_{1/2}^{\otimes 6}    &=& D_3 \oplus 5 D_2 \oplus 9 D_1 \oplus 5 D_0, \ldots, 
\end{align}
\noindent and so on. 
\noindent Since the spin of a quantum state belonging to $D_0$ representation is zero, the action of the energy operator $S^z$ on  such singlet 
state gives zero, and the actions of  the ladder operators $S^+$ and $S^-$ also give zero in total \cite{ZR1997}.

\section{Quantum Reservoir Processing \label{qrc:sec}}
Presence of quantum superposition states is a key element 
that provides superiority of quantum computers over classical 
ones when solving a problem by means of quantum  algorithms.
For a wide class of problems an {\em algorithmic} solution is not 
effective, even in the classical case. The example is image classification, where the task of the processor (either classical, or quantum) is to assign to each input data vector 
an output vector, which labels the class. The dimension of 
input vectors is typically much higher than that of the output. 
A typical example is the classification between cats and dogs, 
where the input size is the total size of all pixels in the image, and the output size is one bit.
For such non-algorithmic problems we have to construct a 
mapping from a high-dimensional space of input data to a 
low-dimensional space of output classes. Since both the input and the output  may be either classical (C) or quantum (Q), there may 
be four different types of data processing: CC, CQ,QC, QQ.

In case both the input and output data are classical, a universal 
function approximator, providing such mapping, is a multilayer 
perceptron 
\begin{equation}
	y^{i+1} = f^{(i)}(\sum_k w^{(i)}_{ks}y^{(i)}_s + b^{(i)}_k),
	\label{fp}
\end{equation}
where $f^{(i)}(\cdot)$ is nonlinear neuron activation function 
on the $i$-th layer of the network. The inputs $y^{(i)}$ are fed 
to this function with a set of tunable weights and biases 
$\{ w^{(i)}_{ks},b^{(i)}_k \}$, which are being optimized during 
the learning process. 
For a given learning set $\{ y^{(0)}(s)\}_s$ with the known set 
of labels $\{ d(s)\}_s$ the learning consists in the optimization 
of weights and biases in a way that they minimize the 
average error $E(\omega,b) = \langle (y^{(N)}(s)-d(s))^2\rangle_s$, or cost function, for some suitable measure on the 
training set \cite{Haykin2009}.

The feed-forward networks, which are widely used in AI systems, 
form only one class of neural networks. In fact, any system comprised of identical units (neurons), generally with all-to-all connections, can be used as a neural network if we manage to 
optimize connection weights in a way that for all input vectors from the training set it output is suitably close to the desired 
values. 

Due to linearity of quantum mechanics there is no straightforward 
way to implement the {\em non-linear} activation function \eqref{fp} in 
quantum settings and fed a linear combination of input states to it. It is, however, possible to harness quantum transitions and 
quantum tunneling to drive the whole quantum system of interconnected qubits (neurons) of a data processing unit to 
a quantum state which minimizes a cost function, and therefore 
provides a solution to a given optimization problem. This 
is the way commercial quantum annealers produced by D-Wave Systems Inc., implemented as Boltzmann machines on SQUIDs, operate \cite{D-wave2011}.  

In classical artificial neural network with all-to-all connections 
the learning procedure requires   tuning of $O(N^2)$ weights. 
This is computationally expensive. The idea of {\em reservoir network} arose from recurrent neural networks \cite{Paquot2012,Larger2017}, 
and consists in tuning only the weights of the output layer. All other weights of a densely connected network -- reservoir -- are set to 
fixed random values at the beginning, and play the role of high-dimensional nonlinear 
kernel, which dramatically increase the capacities of the network, 
if compared to a linear model. Reservoir networks exist in both classical and quantum \cite{Paquot2012,Ghosh2019} versions.

\begin{figure}[t!]\centering
	\includegraphics[width=.49\textwidth]{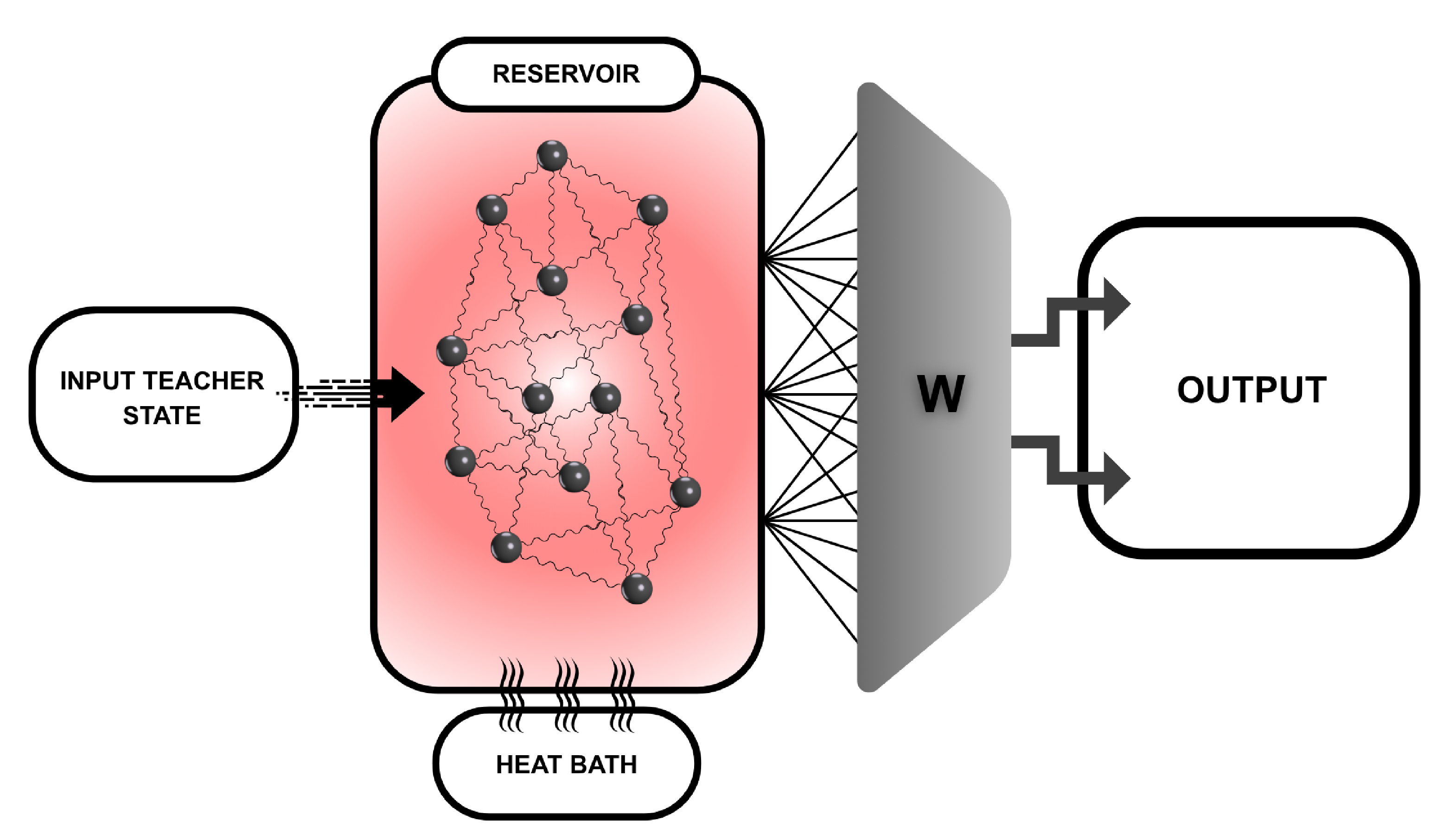}
	\caption{Operation scheme of reservoir computing. Quantum input state is submitted to the input 
		of a complex quantum system of identical particles, interacting to each other as a randomly connected network. This is called 
		''reservoir''. The reservoir incoherently interacts with its environment -- heat bath. The quantum state at the output 
		of the reservoir is filtered by weight matrix $W^{out}$ with tunable weights. The results of this filtering go to  output 
		of the system. Redrawn from \cite{Ghosh2019}.}
	\label{qrc:pic}
\end{figure}

The operation flow chart of reservoir network is schematically 
drawn in Fig.~\ref{qrc:pic}.     
The input data vector is fed to the reservoir -- densely connected network 
with random weights. A small subset of reservoir network, called the 
output layer, is connected to the system output by a set of tunable weights $W^{out}_{ij}$.  Supervised learning process consists 
in tuning this very set of weights in order to minimize the 
learning error on the training set. Since the number of tunable weights is small, the learning process goes fast. 

The generalization of reservoir network to the quantum case is 
straightforward \cite{Ghosh2019}, and specially attractive for the use 
of decoherence-free subspaces \cite{AK2025J,AK2025DD}.

\section{Data classification with quantum reservoir network \label{dc:sec}}

We have simulated the dynamics of six-qubit reservoir learning to distinguish between entangled and product states fed to the input of the 
network. 
The reservoir is described by a Hamiltonian 
\begin{equation}
	H_R = \sum_{i,j=0}^5J_{ij}(b_i^\dagger b_j + b_j^\dagger b_i), \label{hr}
\end{equation}
where $J_{ij}$ are random weights, homogeneously 
distributed in $[-1,1]$ before the beginning of simulation, $b_i^\dagger$ and $b_i$ are the ladder operators of the $i$-th qubit in the reservoir.

The interaction between the reservoir qubits and the teacher quantum system 
is described by interaction Hamiltonian 
\begin{equation}
	H_I(t) = (1-\theta(t-\tau)) \sum_{k=6}^7 \sum_{j=0}^5 f_k W_{j} \left(a_k^\dagger b_j 
	+ b_j^\dagger a_k \right), \label{h7}
\end{equation} 
where $(a_k,a^\dagger_k),k=6,7$ are the ladder operators for the teacher, $W_{j}$ are random input connection weights,
$f_k$ are the coupling constants of the teacher system, which we set to constant values, $\theta(\cdot)$ is the step function, which provides switching the interaction between the reservoir and the teacher on for a finite time $\tau$.

At the initial instant of time the quantum states of the teacher system and the reservoir are independent, and the density matrix $\rho$ of the 
whole system [teacher$+$reservoir] is a direct product 
$$
\rho = \rho^{teach} \otimes \rho^{reservoir}, 
$$
where $\rho^{reservoir}$ is the $64\times64$ density matrix 
of the reservoir qubits.

The evolution  of  total density matrix 
$\rho$ is given by  Lindblad-Gorini-Kossakowski-Sudarshan master equation \cite{Ghosh2019}:
\begin{equation}  
	\imath  \frac{\partial\rho}{\partial t} = [ H_R + H_I,\rho] + \imath \frac{\gamma}{2} \sum_j \cL(b_j) + \imath \frac{P}{2}
	\sum_j \cL(b_j^\dagger),	
	\label{lbd}
\end{equation}
where $H_R$ is the reservoir Hamiltonian \eqref{hr}.
The interaction of reservoir  with the fluctuating environment is described in  
Lindblad-Gorini-Kossakowski-Sudarshan approximation by the dissipation operator 
$$
\cL(x) := 2 x \rho x^\dagger - x^\dagger x \rho - \rho x^\dagger x.
$$
We assume that the pumping ($P$) and the dissipation ($\gamma$) are the same for all 6 qubits of the reservoir.

Before starting the learning process we equilibrate 
the six-qubit reservoir density matrix $\rho^{reservoir}$, 
according to the same master equation \eqref{lbd}, but 
at the absence of any teacher. The equilibration 
can be done either from a fixed [ground] state of the 6 qubit system, or from a random product state 
$\rho^{reservoir}_*= \otimes_{i=0}^5 \rho_i$, with $\rho_i$ being random pure states of individual qubits.
The equilibration of initial state of 6 qubit reservoir in shown in Fig.~\ref{eq6:pic} below.
\begin{figure}[ht]
	\centering \includegraphics[width=8cm]{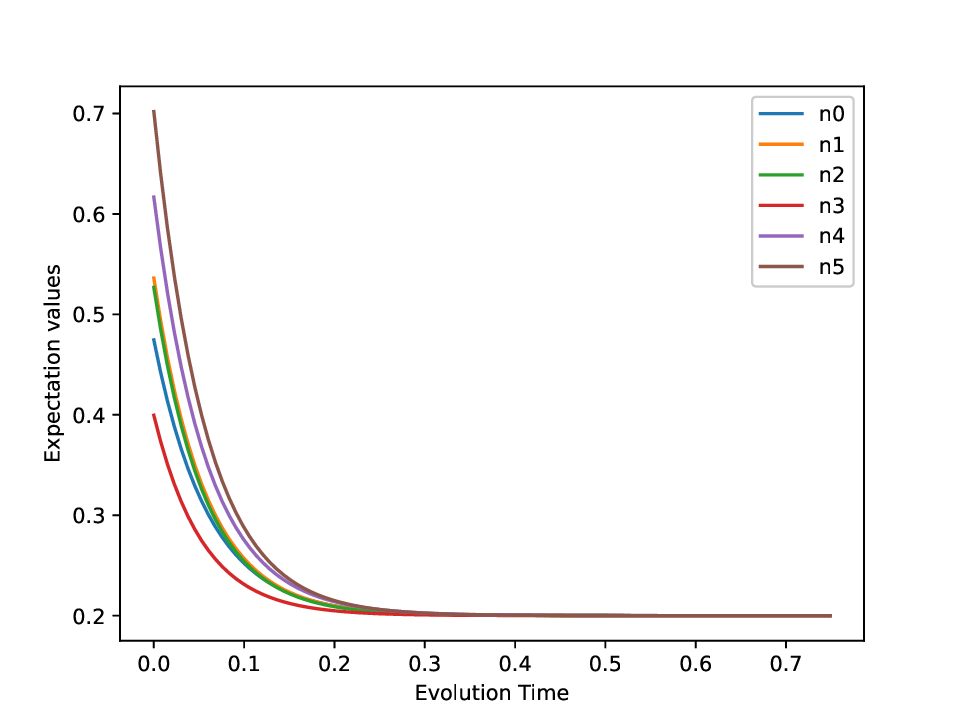}\\
	\includegraphics[width=8cm]{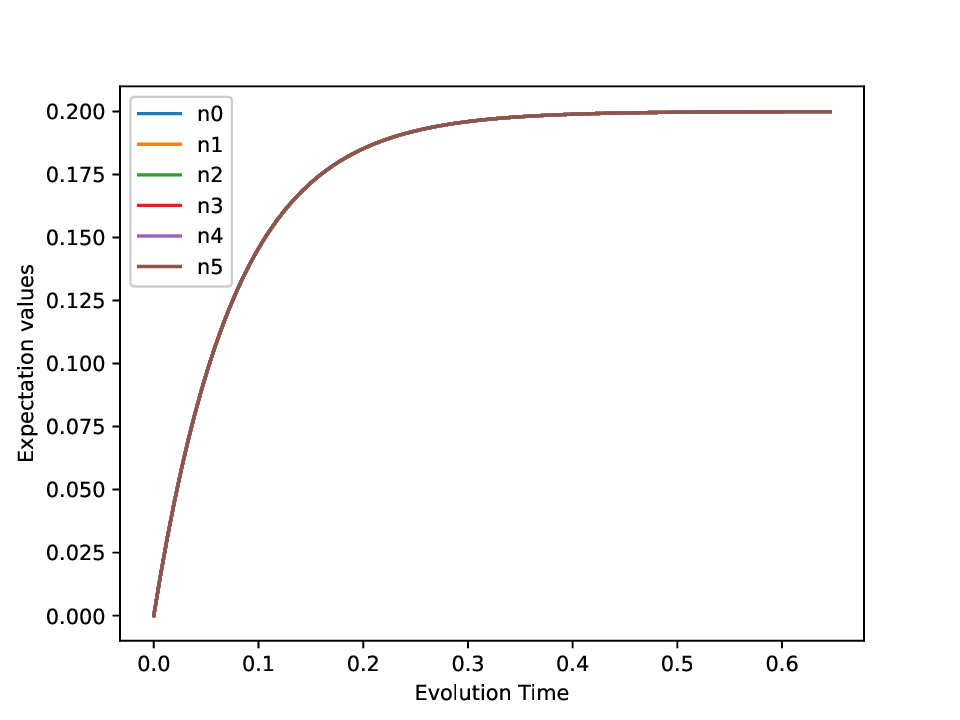}\\
	\caption{Equilibration of ground state of 6 qubit reservoir by Lindblad evolution. In the bottom and top pictures we plot the mean populations $n_i = b_i^\dagger b_i$ for the evolution from a random product state of 6 qubits (top), and from the common ground state (bottom), respectively. 
		In the latter case all populations behave identically.
		Following \cite{Ghosh2019}, time $t$ is measured in units of inverse spectral radius of the reservoir Hamiltonian \eqref{hr}, determined by a set of random coefficients 
		$J_{ij}$}
	\label{eq6:pic}
\end{figure}

The task of the reservoir network is to classify 
the input states $\rho^{teach}$ into two classes: 
the entangled and the separable states. Having submitted 
the input state $\rho^{teach}$ to the reservoir, the state of 
the reservoir is changed due to its interaction with 
the teacher, given by $H_I$ Hamiltonian \eqref{h7}, and the interaction 
with environment governed by master equation \eqref{lbd}. 

Having the interaction between the reservoir and the 
teacher during the time $\tau$, at the instant of the 
read-out of  reservoir state $t_*\ge\tau$, we yield the 
output of reservoir network weighted by a set of 
tunable output weights  $\{ W_j^{out}\}$ in appropriate 
basis. Since we know the labels for the states in the 
training set (1-entangled,0-separable), we can optimize   
the set of weights 
$\{ W_j^{out}\}$ by linear regression providing the 
best output for a given training set.

In our simulations we used two different schemes for preparation of learning samples. The first 
method was to use two separate 
qubits, prepared in either product or an entangled 
state, and coupled to the reservoir by Hamiltonian 
\eqref{h7}. In this method the teacher states were two-qubit states, which were either entangled or 
separable states. In the second method, described 
in \cite{Ghosh2019}, the training set was constructed 
from two-mode squeezed thermal states described in 
Fock basis.

\subsection{Classification of two-qubit states} 
For classification of two-qubit states we used 
standard $\{0,1\}$ qubit basis. The teacher states were 
given by $4\times4$ density matrices  $\rho^{teach}$ 
defined as follows:
\begin{description}
	\item[Product states] with density matrices generated 
	as direct product of two one-qubit states 
	$$\rho_{separ}^{teach} = |i\ket \rho^1_{ij}\bra j| \otimes 
	|k\ket \rho^2_{kl}\bra l|
	$$
	\item[Entangled states] General $4\times4$  Hermitian matrices 
	$\rho^{teach}$ with unit trace, constrained from below by the threshold minimal value 
	of log-negativity [$\nu_{min}=0.15$ in our simulations]. 
\end{description}

The learning process was organized as follows. Having set the random input weights $W_j, j=\overline{0,5}$ and the reservoir Hamiltonian couplings $J_{i,j}, j,j=\overline{0,5}$ parameters, the state of six-qubit  reservoir was equilibrated according to the master 
equation \eqref{lbd} during the time $T_{equilibration}$. 
In learning cycle we sequentially generate random two-qubit teacher states, as defined above, and  connect them to  reservoir qubits by the Hamiltonian \eqref{h7} during the time $T_{reading}$. 

The process of mutual evolution of the total $6+2$ qubit system ('reading' process) was implemented by Lindblad-Gorini-Kossakowski-Sudarshan master equation \eqref{lbd} using standard {\sl qutip Python library} for  open quantum system simulations \cite{qutip5}.

The novelty of our approach (if compared to \cite{Ghosh2019}) is the reading of 5d output 
performed by taking the expectation values of five projectors onto 
5 singlet states appearing in the direct product of six qubit states \cite{ZR1997}
\begin{equation}
	P_I = |I\rangle \langle I|.
\end{equation} 
These expectation values are given by the traces of convolution of projection operators with the reservoir 
density matrix $\rho^{reservoir}$. For technical reasons 
the numerical simulation was accomplished in 8 qubit basis with the total density matrix $\rho$, followed by tracing over the states of teacher qubits ($i=6,7)$, so that 
\begin{equation}
	m_I = \tr (\rho P_I) 
	\label{m5}
\end{equation}
Five expectation values $m_I$ were fed to the linear regression 
\begin{equation}
	\sum_{I=1}^5 W^{out}_I m_I(X) = Y(X),
\end{equation} 
where $Y(X)\in \{0,1\}$ is the label of the class for the 
element $X$ of the training set. Linear regression was provided by {\sl sklearn Python library} \cite{scikit}.

\begin{figure}[ht]
	\centering \includegraphics[width=8cm]{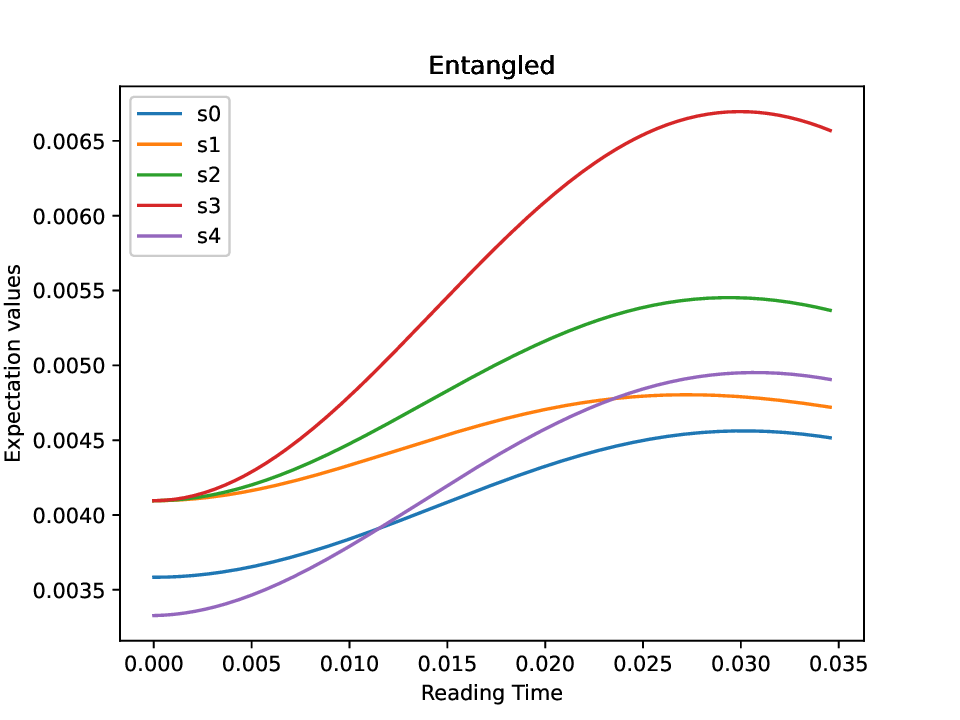}\\
	\centering \includegraphics[width=8cm]{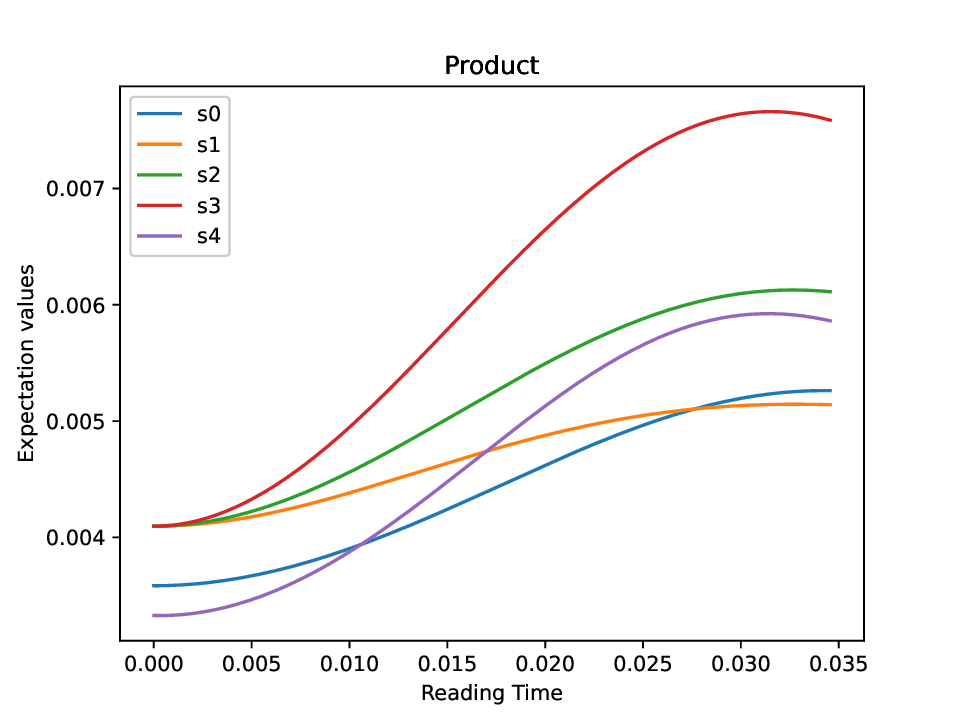}
	\caption{Typical Lindblad evolution of 5 singlet expectation values 
		$m_0,\ldots,m_4$ during the 'reading ' process by interaction of 6 qubit reservoir with 2 teacher qubits. 
		The evolution is shown for an entangled teacher state (top), 
		and product teacher state (bottom)
	}
	\label{e5:pic}
\end{figure}

A typical Lindblad evolution of five singlet expectation values \eqref{m5} during the 'reading' process is shown in Fig.~\ref{e5:pic}, for entangled (top) and the product (bottom) teacher states, respectively.
The parameters used in our simulations of the learning process were
\begin{align}\nonumber 
	\gamma = \max |\mathrm{Eigenvalues}(H_R)|,  P = 0.5\gamma \\
	T_{equilibration} = 0.2\gamma,   T_{reading} = 0.01\gamma,
	\label{par:def} 	
\end{align} 
that is all time intervals were measured in the units of inverse spectral radius of Hamiltonian $H_R$. The teachers coupling constants were set to big values 
$f_k=10.0$ providing the strong entanglement process between the teacher and the 
reservoir qubits.

\begin{figure}[ht]
	\centering\includegraphics[width=8cm]{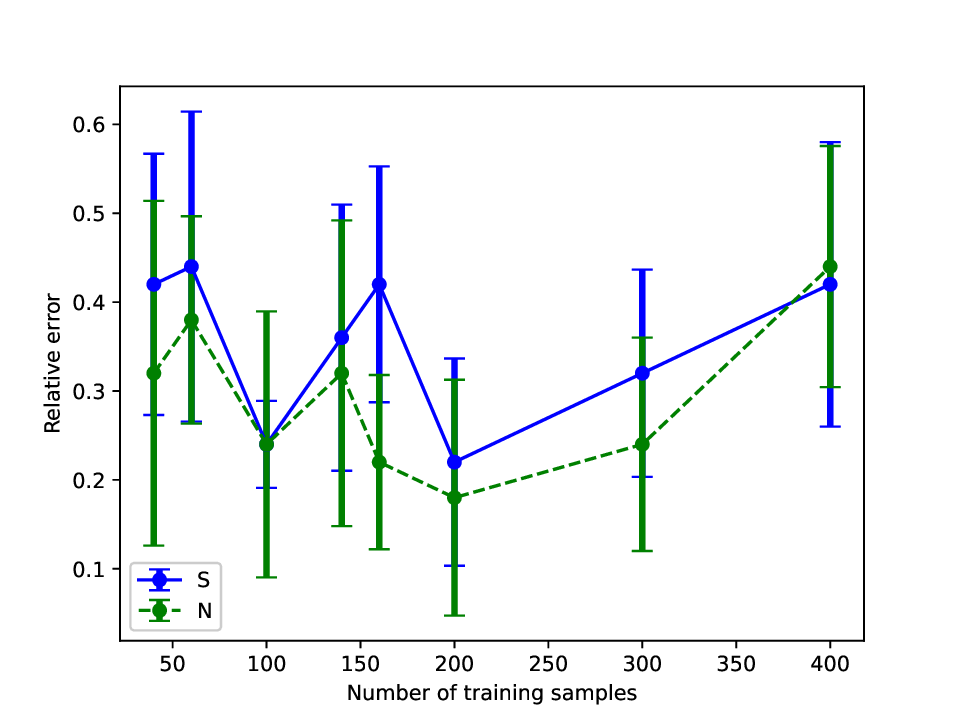}
	\caption{Comparison of reservoir learning processes performed in our 5 singlet basis (S) and that performed 
		in 6 population basis (N) according to \cite{Ghosh2019}}
	\label{qubit-learn:pic}
\end{figure}

The results of tests of the trained network on the 
training set of 10 random two-qubit samples  
are presented in Fig.~\ref{qubit-learn:pic}
We compare the performance of our singlet-based reservoir 
network with that of quantum reservoir network which 
uses mean populations $\langle n_i\rangle=\langle b_i^\dagger b_i\rangle$ of all six qubits $i=\overline{0,5}$ for the output \cite{Ghosh2019}. As it can be seen from Fig.~\ref{qubit-learn:pic}, the performance of such reservoir network is rather modest for either of cases: (S) our singlet basis, (N) population basis. The relative recognition error is at best 
in $\epsilon \in [0.2,0.3]$ for a small number of training samples in a range of few hundreds  of samples. Training sets of larger size drive reservoir network to over-learning events.
The reason for such performance is a relatively small number 
of degrees of freedom of two-qubit teacher that are imprinted 
in the reservoir dynamics during the 'reading' process. 

In the next section we have significantly improved 
the performance of reservoir learning by choosing two-mode squeezed thermal states as a training set, as was implemented 
in \cite{Ghosh2019}. Even for a small dimension of the 
Fock state, $N_F=4$, the teacher state density matrix 
for squeezed states becomes $16\times16$ density matrix (rather than $4\times4$ density matrix for two-qubit states). This 
makes the learning process essentially more effective.

\subsection{Classification of two-mode squeezed states}
In the second method, following \cite{Ghosh2019}, 
we used two-mode Fock basis, with maximal mode occupation number $N_F-1=3$. This gives $16\times16$ density matrix $\rho^{teach}$ for the input states. In the case of {\em product states} this matrix is generated as a direct product of two independent $4\times4$ density matrices, 
of the first and second mode, respectively. 
For the {\em entangled states} the density matrix is generated by means of squeezing of thermal states.

The mean occupation number of a boson mode in the thermal state is defined as 
\begin{equation}
	\bar n = \frac{1}{Z(\beta)} \sum_{n=0}^\infty n e^{-\beta n}, \quad Z(\beta) = \sum_{n=0}^\infty  e^{-\beta n},
\end{equation}
which gives 
\begin{equation}
	\bar{n} = \frac{1}{e^\beta-1}, \quad \beta = \ln \left(1+\frac{1}{\bar{n}}\right).
\end{equation}
If the total occupation number is restricted by 
the maximal occupation number $N_F-1$, the probability 
of having $n_1$ quanta in the first mode and 
$n_2$ quanta in the second mode, is given by 
\begin{equation}
	\rho_{n_1,n_2} = \frac{e^{-\beta(n_1+n_2)}}{Z_1^2(\beta)},\quad 
	Z_1(\beta)= \sum_{j=0}^{N_F-1} e^{-\beta j} \label{rprob}
\end{equation}
The latter equation is symmetric with respect to first and second modes and depends on the sum of populations only. The density matrix of a totally 
mixed thermal state is 
\begin{equation}
	\hat{\rho}_{th} = \sum_{n_1,n_2=0}^{N_F-1} 
	|n_1,n_2\rangle \rho_{n_1,n_2} \langle n_1,n_2|.
	\label{r25}
\end{equation}
The action of annihilation and creation operators 
(of each mode) in the Fock basis are determined as 
$$
a|n\rangle = \sqrt{n}|n-1\rangle, \quad 
a^\dagger|n\rangle = \sqrt{n+1}|n+1\rangle.
$$ 
So, that for $N_F=4$ the single-mode annihilation operator is given by $4\times4$ matrix 
$$
a = \begin{pmatrix}
	0 & \sqrt{1} & 0 & 0  \cr 
	0 & 0 & \sqrt{2} & 0 \cr 
	0 & 0 & 0 & \sqrt{3} \cr
	0 & 0 & 0 & 0  
\end{pmatrix}.
$$
For the two-mode system, described by $16\times16$ density matrix in $|n_1,n_2\rangle$ Fock basis, there are two annihilation operators 
\begin{equation}
	A_1 = a \otimes \I, \quad A_2 = \I \otimes a,
\end{equation}
for the first and the second mode, respectively, 
and two creation operators $A_1^\dagger,A_2^\dagger$. If the modes are independent, the ladder operators of different modes commute to each other.

The correlated state can be obtained from a thermal 
two-mode state \eqref{r25} by applying  {\em squeezing} operator
\begin{equation}
	S = \exp \left(\alpha A_1^\dagger A_2^\dagger 
	- \bar{\alpha} A_1 A_2 \right), \label{sqo}
\end{equation}
where $\alpha$ is arbitrary complex number, 
and $\bar{\alpha}$ denotes its complex conjugation.

For arbitrary complex number $\alpha$, applying  operator $S$, we can map a mixed thermal state 
\eqref{r25} into a squeezed state 
\begin{equation}
	\hat{\rho}_{th} \to \upsilon = S \hat{\rho}_{th} S^\dagger .\label{sqs}
\end{equation} 
For simulation of squeezed states $\upsilon$ we went along the lines 
of the work \cite{Ghosh2019}. To do so we used random parameters uniformly distributed in the domains 
\begin{equation}
	\phi \in \bigl[\frac{1}{2}-\frac{\pi}{10},
	\frac{1}{2}+\frac{\pi}{10}
	\bigr], s \in [0.8,0.95], \theta \in [0,2\pi]
	\label{sqpar}
\end{equation}
with these parameters we have defined the mean occupation number 
of two modes $\bar{n}$ and the inverse temperature $\beta$ as 
\begin{equation}
	\bar{n}=s^2 \cos^2\phi,\quad \beta =\ln \left(1+\frac{1}{\bar{n}}\right).
\end{equation}
Having defined the inverse temperature $\beta$, we construct 
$N^2_F\times N^2_F$ diagonal density matrix for thermal states 
defined by probabilities \eqref{rprob}:
\begin{equation}
	\rho^{thermal}_{n_1 n_2,n_1 n_2} = \mathrm{diag}
	\frac{e^{-\beta (n_1+n_2)}}{Z_1^2}. \label{tstate}
\end{equation}
To generate entangled two-mode states we squeeze thermal states 
\eqref{tstate} with squeezing parameters uniformly distributed 
in the domain \eqref{sqpar}. Namely, the complex squeezing parameter 
is given by 
\begin{equation}
	\alpha = s \sin(\phi)e^{\imath\theta},
\end{equation} 
so that  squeezing operator is given by (\ref{sqo},\ref{sqs}).
For each random sample of parameters for 
the squeezed state $\upsilon$ we calculated  
the logarithmic negativity to estimate how much the 
state is entangled. In our simulation we chose threshold 
$\nu_{min}=0.15$, above which we consider the simulated state 
$\upsilon$ as entangled. 
The logarithmic negativity is calculated in a usual way \cite{VidalWerner2002,Plenio05}. We take matrix $G$, which is a partial transpose of matrix $\upsilon$ with respect to the indices of the second mode. The logarithmic negativity is then given by 
\begin{equation}
	\mathrm{logneg}(\upsilon) = \log_2 \Tr (GG^\dagger).
\end{equation}

We taught the reservoir network by even number of samples, where 
each odd sample was a product state given by a direct product of 
two $N_F\times N_F$ random density matrices, and each even sample taken as a squeezed state \eqref{sqs}, with a log-negativity exceeding the threshold value $\nu_{min}$. In the simulation with Fock state density matrices we used the same set of evolution parameters \eqref{par:def} and coupling constants as for two-qubit teacher simulations. However, since the density matrix for two-mode system with $N=4$ is 16 times larger that for two-qubit system, the learning process turns to be more effective.

\begin{figure}[ht]
	\centering \includegraphics[width=8cm]{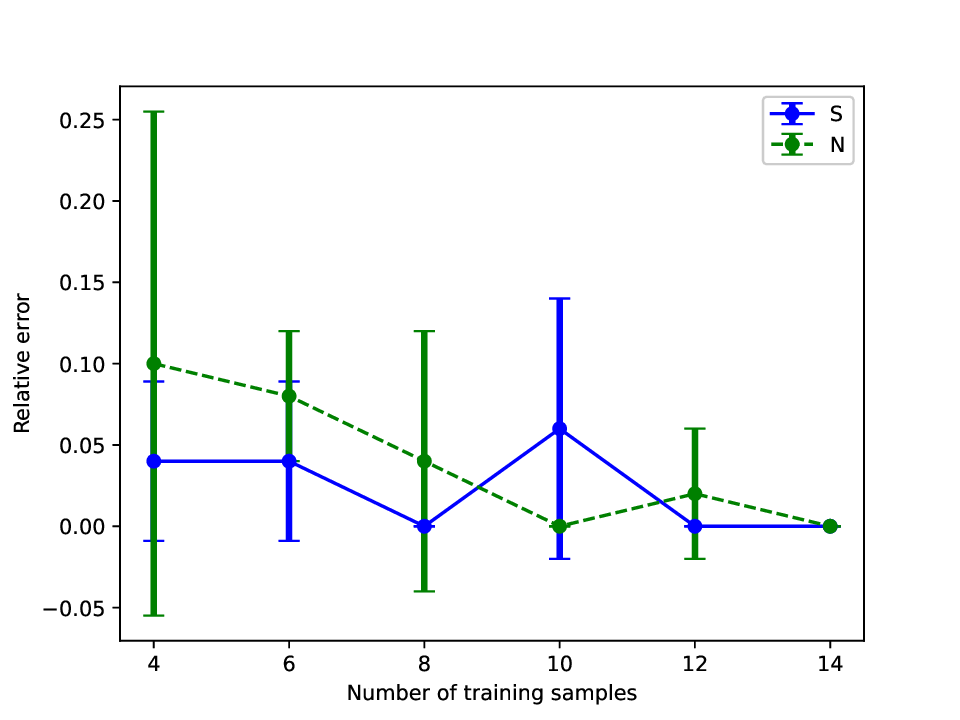}
	\caption{Comparison of the relative classification errors measured on a 10 sample test set for quantum reservoir learning for different volumes of the training set. The graphs are shown for singlet-based network (S) and the population based network (N)}
	\label{fock-err:pic}
\end{figure}

As we have observed, the learning process effectively starts from a 4 sample set, which contains only two entangled and two product states of two modes in Fock basis, see Fig.~\ref{fock-err:pic}.
Starting from the training sets of approximately 12 learning samples, the recognition on a 12 sample test set becomes almost ideal: no errors were observed during 5 runs on the same volume set.  
We have observed recognition errors on the 10 sample test set 
only when the volume of the training set was less than 14, otherwise, all 10 samples received correct recognition during at least 5 runs. We were not able to observe any over-learning effects for the training set volume below 3000 samples.

 \section{Physical readout of the singlets}\label{readout:sec}
 
 In the original decoherence-free-subspace construction of Zanardi and Rasetti \cite{ZR1997}, the encoding principle was specified without a corresponding physical protocol for reading out the individual singlet components. The distinction between the protected evolution and the final readout is important. In the Dicke limit \cite{Dicke1954-yh}, the environment couples identically to the qubits and cannot distinguish their individual labels. The system--bath interaction can therefore be expressed in terms of collective spin operators $S_\alpha$ \cite{Kempe2001-xq,Quiroz2024-jl}. 
 
 A measurement that distinguishes the five singlets must resolve how the individual qubits are paired. Consequently, a readout based on single-qubit operations cannot preserve the indistinguishability of the qubits, on which the collective-noise protection relies during the reservoir evolution. These requirements are nevertheless compatible when they are assigned to different stages of the experiment: the collective symmetry is maintained during the reservoir evolution, whereas it is deliberately broken only at the end of a reading cycle. The final measurement is destructive, but the corresponding output features can be estimated from repeated preparations of the same input and reservoir state. This is the sense in which the following protocol is semi-qualitative: it establishes a physically meaningful feature map without attempting to predict the complete experimental classification error.
 
 For notational consistency with the usual pair construction, we label the six qubits locally by $1,\ldots,6$ in this section; these labels correspond to the three pairs $(0,1)$, $(2,3)$, and $(4,5)$ in the zero-based indexing used elsewhere in the manuscript. For two qubits, the swap operator is \cite{Kempe2001-xq}
 \begin{equation}
 	\Pi_{ij}=\frac{1}{2}\left(\mathbb{I}+\boldsymbol{\sigma}_i\cdot\boldsymbol{\sigma}_j\right),
 	\label{swap:operator}
 \end{equation}
 which equals $-1$ on the pair singlet and $+1$ on the pair triplet. More generally, the squared total spin $\mathbf{S}_k^2$ of a set of $k$ spin-$1/2$ qubits (spin operators $\boldsymbol\sigma/2$, so that $\mathbf S^2=J(J+1)$) can be written as \cite{Kempe2001-xq}
 \begin{equation}
 	\mathbf{S}_{k}^{2}
 	=k\left(1-\frac{k}{4}\right)\mathbb{I}
 	+\frac{1}{2}\sum_{i\neq j}\Pi_{ij}.
 	\label{total:spin}
 \end{equation}
 For an individual pair this gives (derived here from Eq.~\eqref{total:spin} with $k=2$)
 \begin{equation}
 	\mathbf{S}_{ij}^{2}=\Pi_{ij}+\mathbb{I},
 	\label{pair:spin}
 \end{equation}
 whose eigenvalue is $0$ for a pair singlet and $2$ for a pair triplet. Since exchange operators belong to the commutant of the collective $SU(2)$ action \cite{Kempe2001-xq}, the three pair-spin operators commute with the collective generators. They also commute with one another because they act on disjoint pairs.
 
 We label each pair by $s_{ij}=0$ for a singlet and $s_{ij}=1$ for a triplet, and consider the pairing $(12)$, $(34)$, and $(56)$. Coupling the three pair spins to total spin zero gives the five labels (by angular-momentum addition of the pair spins; checked here by explicit construction)
 \begin{equation}
 	\begin{aligned}
 		\mathcal{A}=\{ &(0,0,0),(0,1,1),(1,0,1),\\
 		&(1,1,0),(1,1,1)\}.
 	\end{aligned}
 	\label{allowed:labels}
 \end{equation}
 Each label in $\mathcal{A}$ identifies one singlet vector, and hence defines a pair-coupling basis $\{|I\rangle\}_{I=1}^{5}$ of the six-qubit singlet sector (only $|I\rangle\leftrightarrow(0,0,0)$ is a product of pair singlets; the other four are pair triplets coupled to total spin zero). This basis must be specified when comparing a physical readout with the projectors used in the numerical model. If the projectors $P_I=|I\rangle\langle I|$ are defined using a different pairing or coupling order, the two sets of singlet states are related by a $5\times5$ unitary (a recoupling transformation). The sum $\sum_I P_I$, the projector onto the $5D_0$ block, does not depend on this choice, but the individual populations $\operatorname{Tr}(\rho P_I)$ do, because they depend on the coherences of $\rho$ within the singlet block.
 
 The pair labels can be measured using a Bell-basis measurement \cite{Bennett1993-dp}. For each pair, a CNOT is applied with the first qubit as the control, followed by a Hadamard gate on the control and computational-basis measurements of both qubits \cite{NielsenChuang2000}
 With this convention, the singlet is mapped to the bit string $(1,1)$, while the three triplet states are mapped to $(0,1)$, $(1,0)$, and $(0,0)$. The three pair measurements act on disjoint qubits and can therefore be executed in parallel, without ancillary qubits and with constant circuit depth. Each experimental shot produces one of the eight label strings
 \begin{equation}
 	\mathbf{s}=(s_{12},s_{34},s_{56}),
 \end{equation}
 and the corresponding frequencies are estimated as
 \begin{equation}
 	f_{\mathbf{s}}=\frac{n_{\mathbf{s}}}{N_{\rm shots}},
 \end{equation}
 where $n_{\mathbf{s}}$ is the number of occurrences of the label $\mathbf{s}$. For an individual marginal frequency, the finite-shot standard error has the usual binomial scaling $\sqrt{f(1-f)/N_{\rm shots}}$, although the complete set of eight frequencies is multinomially correlated because their sum is one.
 
 Let $Q_{\mathbf{s}}$ denote the joint projector onto the three pair labels $\mathbf{s}$, i.e.\ onto the six-qubit states in which the pairs have exactly the spins $\mathbf s$; $\operatorname{Tr}(\rho Q_{\mathbf s})$ is the probability of the outcome $\mathbf s$. For an allowed label $\mathbf{s}(I)\in\mathcal{A}$, the corresponding physical feature is
 \begin{equation}
 	\begin{split}
 		f_I &= \operatorname{Tr}\!\bigl(\rho\,Q_{\mathbf s(I)}\bigr)
 		= \operatorname{Tr}(\rho P_I)+\lambda_I,\\
 		\lambda_I &= \operatorname{Tr}\!\bigl[\rho\,(Q_{\mathbf s(I)}-P_I)\bigr]\ge 0 .
 	\end{split}
 \end{equation}
 The quantity $\lambda_I$ is the contribution from states outside the singlet sector that possess the same pair labels. Therefore, the experimentally measured frequency is equal to the ideal numerical feature $m_I=\operatorname{Tr}(\rho P_I)$ only when the reservoir state is confined to the singlet sector and the same pair-coupling basis is used. In particular, no state-independent numerical value can be assigned to $\lambda_I$; it is determined by the leakage of the actual reservoir state. For the label $(0,0,0)$, the pair-label sector contains only the singlet contribution, so $\lambda_I=0$ identically. For all five labels, $\lambda_I$ vanishes when $\rho$ has support entirely in the singlet sector.
 
 The three outcomes outside $\mathcal{A}$,
 \begin{equation}
 	(0,0,1),\qquad (0,1,0),\qquad (1,0,0),
 \end{equation}
 can be accumulated into the diagnostic frequency
 \begin{equation}
 	f_{\rm leak}=\sum_{\mathbf{s}\notin\mathcal{A}}f_{\mathbf{s}}.
 	\label{leakage:frequency}
 \end{equation}
 The six-qubit decomposition contains total-spin sectors $J=0,1,2,3$ with multiplicities $5,9,5,1$, respectively \cite{Kempe2001-xq}. The label $(0,0,0)$ is associated only with $J=0$; the labels $(0,1,1)$, $(1,0,1)$, and $(1,1,0)$ may occur for $J=0,1,2$; and $(1,1,1)$ may occur for $J=0,1,2,3$ (angular-momentum addition of the pair spins, checked here by explicit construction). The three labels entering Eq.~\eqref{leakage:frequency} occur only in the $J=1$ sector. They therefore certify a particular type of leakage, but not leakage in general: only three of the nine $J=1$ multiplets are flagged, while the $J=2$ and $J=3$ sectors are not detected by this diagnostic. The five frequencies $f_I$ can consequently be used as the features of the linear output layer, with $f_{\rm leak}$ retained as an auxiliary quality indicator rather than as a complete leakage estimator.
 
 The invariance of the label observables under collective dissipation follows from their commutation with the collective generators. For Markovian noise in Lindblad form \cite{Lindblad1976-mv}, \cite{Gorini1976-qv}, \cite{Kempe2001-xq} with jump operators constructed from $S_\alpha$ (or equivalently from $S_\pm$ and $S_z$), the adjoint dissipator satisfies
 \begin{equation}
 	\mathcal{D}^{\dagger}(Q_{\mathbf{s}})
 	=\sum_k\gamma_k\left(L_k^{\dagger}Q_{\mathbf{s}}L_k
 	-\frac{1}{2}\left\{L_k^{\dagger}L_k,Q_{\mathbf{s}}\right\}\right)=0,
 	\label{adjoint:dissipator}
 \end{equation}
 provided that $[Q_{\mathbf{s}},L_k]=[Q_{\mathbf{s}},L_k^{\dagger}]=0$ (derived here; checked numerically for $L_k\in\{S_-,S_+,S_z\}$). This statement concerns the ideal collective-noise model only; local noise acting differently on individual qubits is not removed by the DFS construction, and the Hamiltonian part of the evolution must also be compatible with the DFS \cite{Kempe2001-xq}.
 
 The readout also requires that the reservoir dynamics do not appreciably change the pair labels during the measurement sequence. This can be ensured by switching off the relevant inter-pair couplings or by making the readout time short compared with the corresponding interaction time. Thus, collective-noise protection during the reservoir evolution and faithful pair-label measurement during the final readout are separate requirements. The local gates used for the Bell measurement break the collective symmetry and introduce ordinary gate and assignment errors, while the destructive measurement requires a new state preparation for each shot. A quantitative treatment of these errors, as well as platform-specific singlet--triplet detection schemes such as Pauli spin blockade with resonator detection, for which single-shot readout of a two-electron spin state has been demonstrated in silicon \cite{Zheng2019-ox}, is beyond the scope of the present semi-qualitative discussion. The protocol nevertheless provides a concrete physical route to approximate the ideal projections used by the reservoir model by experimentally accessible frequency features $f_I$, with the bias $\lambda_I$ and the basis choice stated above.
 
 A proper gate-based realization of the current work would require significant effort. The circuit depth and two-qubit-gate overhead may become significant in such contexts. A topology-driven quantum architecture and approximate brick-wall compilation can mitigate these problems to some extent \cite{Su2025-it}. However, a detailed implementation plan and associated noise analysis are beyond the scope of this letter and are left for future work.

\section{Conclusion}
Classical reservoir networks have emerged from recurrent 
neural networks as an effective tool for analysis of the 
time-dependent data. This method is believed to mimic in 
some way the operations of biological brain \cite{Larger2017}.
Quantum reservoir networks are of their own interest for classification of quantum data. They are also considered as 
alternative to quantum state tomography \cite{Zia2025}.

In this paper we consider quantum reservoir networks 
from the standpoint whether at certain conditions they 
can perform quantum information processing at ambient temperature, without a special cooling system as that 
demanded by common quantum computers. The presented 
mathematical toy-model harnesses the reading of quantum 
reservoir state by means of its projections onto decoherence-free subspace. These projections are not affected by large-scale environmental fluctuations, provided that the environment couples collectively to all qubits.

Zanardi and Rasetti \cite{ZR1997} did not specify how the singlet states amplitudes are to be read out. A physical readout based on Bell-basis measurements of three qubit pairs is proposed in Sec.~\ref{readout:sec}; it gives pair-label frequencies that approximate the singlet projections, and its gate and assignment errors and the leakage out of the singlet sector remain to be assessed quantitatively. However, since in the theoretical model the amplitudes can 
survive under collective environmental fluctuations, a reservoir network with an output in DFS can be considered as a prototype 
of a quantum information processing system with relaxed cooling requirements, provided that a readout compatible with the symmetry can be realized.

On the other hand, since    the classification of quantum 
states into entangled and separable states is of great 
practical importance for quantum communication and quantum 
information processing, we can also think of {\em in silico} 
implementation of such quantum reservoir networks complementary to the known optical reservoir processing 
\cite{Paquot2012,Zia2025}. 
\section*{Acknowledgement}
The authors are thankful to Profs. L.Fedichkin and R.Nazmitdinov for useful comments.

\end{document}